\documentclass{article}

\usepackage{graphicx}
\usepackage{times}
\usepackage{xcolor}
\usepackage{amsfonts}
\usepackage{esvect}
\usepackage{harpoon}
\usepackage{amsmath,accents}

\usepackage{subfigure,color}
\usepackage{cite}

\def\e{\begin{equation}}
\def\f{\end{equation}}
\def\_#1{{\bf #1}}
\def\=#1{\overline{\overline #1}}
\def\/#1{_{\rm #1}}
\def\.{\cdot}

\title{Time Interfaces in Bianisotropic Media}

\newenvironment{sciabstract}{%
\begin{quote} \bf}
{\end{quote}}

\author
{
M.~S.~Mirmoosa$^{1}$\footnote{These authors contributed equally}, M.~H.~Mostafa$^{2}$\footnotemark[1]\,\,\footnote{Correspondence: mohamed.mostafa@aalto.fi}, A.~Norrman$^{1}$, and S.~A.~Tretyakov$^{2}$\\
\normalsize{$^{1}$Department of Physics and Mathematics, University of Eastern Finland, P.O.~Box 111, FI-80101 Joensuu, Finland}\\
\normalsize{$^{2}$Department of Electronics and Nanoengineering, Aalto University, P.O.~Box 15500, FI-00076 Aalto, Finland}
}

\date{}

\begin{document} 

\baselineskip24pt

\maketitle 

%%%%%%%%%%%%%%%%%%%%%%%%%%%%%%%%%%%%%%%%%%%%%%%%%%%%%%%%%%%%%%%%%%%%%%%%%%%%%%%

\begin{sciabstract}
{Wave phenomena in bianisotropic media have been broadly scrutinized in classical electrodynamics, as these media offer additional degrees of freedom to engineer electromagnetic waves. However, all investigations concerning such systems have so far been limited to stationary (time-invariant) media. Temporally varying the magnetoelectric coupling manifesting bianisotropy engenders a unique prospect to manipulate wave-matter interactions in new ways. In this paper, we theoretically contemplate electromagnetic effects in weakly dispersive bianisotropic media of all classes when the corresponding magnetoelectric coupling parameter suddenly jumps in time, creating a time interface in spatially uniform bianisotropic media. We investigate scattering effects at such time interfaces, revealing novel polarization- and direction-dependent phenomena. We anticipate that our work paves the road for further exploration of time-varying bianisotropic metamaterials (metasurfaces) and bianisotropic photonic time crystals, thus opening up interesting possibilities to control wave polarization and amplitude in reciprocal and nonreciprocal manners.}
\end{sciabstract} 

%%%%%%%%%%%%%%%%%%%%%%%%%%%%%%%%%%%%%%%%%%%%%%%%%%%%%%%%%%%%%%%%%%%%%%%%%%%%%%%

\maketitle

%%%%%%%%%%%%%%%%%%%%%%%%%%%%%%%%%%%%%%%%%%%%   Introduction   %%%%%%%%%%%%%%%%%%%%%%%%%%%%%%%%%%%%%%%%%%%%

\section{Introduction}  
Interaction of waves with systems whose effective properties change in time, although remaining uniform in space, has engrossed significant curiosity~\cite{Engheta20NPH,galiffi2022photonics, Ptitcyn2023Tutorial}. In particular, the effects arising from 
rapid changes in material properties in time have been actively studied. One enticing possibility provided due to such temporal discontinuities is the creation of reflected (backward) waves and angular frequency conversion~\cite{Agrawal2014RTC,mendoncca2002time,morgenthaler1958velocity}. For media without magnetoelectric coupling, these phenomena were examined theoretically~\cite{morgenthaler1958velocity,mendoncca2002time,Wilks1988TH} and also confirmed experimentally, for example, in the context of plasma physics~\cite{Yugami2002EXPER,nishida2012EXP}, using water waves~\cite{water}, and in transmission lines operating at megahertz frequencies~\cite{Alu_TL}. Based on the two fundamental phenomena of angular frequency conversion and wave scattering, the notion of photonic time crystals formed by periodically varying media in time was introduced~\cite{PTC_1966,Halevi9PTC,Segev8PTC,WaMIR23PTC}, and, specifically, periodical sharp temporal discontinuities were inspected~\cite{Segev8PTC}. Furthermore, a multitude of effects have been reported including anti-reflection temporal coatings~\cite{ramaccia2020light,pacheco2020anti}, temporal beam
splitting~\cite{mendoncca2003temporal}, inverse prism~\cite{akbarzadeh2018inverse}, temporal aiming~\cite{pacheco2020aiming}, polarization conversion~\cite{xu2021complete}, wave freezing and melting~\cite{wang2023controlling}, as well as transformation of surface waves into free-space radiation~\cite{Grap19SPP,wang2023controlling}. All these findings were mainly associated with investigations of temporal discontinuities in unbounded isotropic or anisotropic bulk media and conventional metasurfaces or sheets. However, to the best of our knowledge, wave phenomena at temporal discontinuities of bianisotropic media or bianisotropic metasurfaces have not been scrutinized, with a single exception of isotropic chiral media \cite{mostafa2023spin,Alu_copy}. 

Bianisotropic media are electromagnetic linear media that exhibit magnetoelectric coupling. From the material relations point of view, this means that the electric and magnetic polarization densities are connected to both the electric and magnetic fields (see,~e.g.,~Ref.~\cite{Serdyukov}). Investigations of these media have a long history which goes back to the studies of optical activity in crystals in the early 19th century (by Arago, Biot, and others)~\cite{Lindell}. Optical activity, and more generally, reciprocal magnetoelectric coupling phenomena were shown to be manifestations of the first-order spatial dispersion in the medium. On the other hand, in the first half of the 20th century, it was found that magnetoelectric coupling can be caused also by some nonreciprocal effects (by Dzyaloshinskii, Astrov, and others). Thanks to this knowledge and also to the considerable research on electromagnetics of moving media, the general concept of bianisotropic media was eventually introduced in 1968 in the electromagnetics parlance~\cite{Kong1,Kong2}. The propagation of electromagnetic waves in such media was overwhelmingly contemplated in the late 20th century (see the review in Ref.~\cite{Serdyukov}), and, subsequently, related studies have been extended to hitherto bianisotropic metasurfaces~(see the review in Ref.~\cite{Nanophotonics_review}). However, the nearly two hundred years of research outlined above was limited primarily to time-invariant bianisotropic systems. It is therefore expected that the exploration of phenomena in time-varying bianisotropic media and metasurfaces will reveal a plurality of new effects and functionalities, complementing the application potentials of stationary bianisotropic materials, metamaterials, and metasurfaces.

In this paper, we make initial steps in this research direction and contemplate the electromagnetic scattering from temporal interfaces between an isotropic magnetodielectric medium and bianisotropic media of all known classes (chiral, Tellegen, moving, and omega \cite{Serdyukov}). For each class, we deduce the scattered fields created at temporal interfaces and explain the corresponding wave phenomena. In particular, we find that chiral and Tellegen temporal interfaces offer an opportunity to control wave polarization in reflection and transmission, while artificial moving and omega temporal interfaces manipulate wave features based on the propagation direction of the incident wave. Also, we list all such possible wave effects at both spatial and temporal interfaces and show that phenomena at temporal interfaces complement those at spatial interfaces, adding possibilities to control not only the waves' amplitude, phase, and polarization, but also their angular frequency. 

The paper is organized as follows: Section~\ref{sec:BMICLTI} succinctly explains the classification of linear time-invariant bianisotropic media and concisely describes the electromagnetic phenomena due to the presence of a spatial interface or a two-dimensional array of bianisotropic inclusions in space. Afterward, Section~\ref{sec:TIBMA4} comprehensively analyses temporal interfaces and the associated wave phenomena for each class of bianisotropic media. Additionally, the same section throws light on several topics including time-domain material relations, field evaporation, and possible realizations of such systems. Finally, Section~\ref{sec:CONCL} summarizes the main conclusions of the paper.

%%%%%%%%%%%%%%%%%%%%%%%%%%%%%%%%%%%%%%%%%%%%%%%%%%%%%%%%%   Section 2   %%%%%%%%%%%%%%%%%%%%%%%%%%%%%%%%%%%%%%%%%%%%%%%%%%%%%%%%%

\section{Bianisotropic Media} 
\label{sec:BMICLTI}

\subsection{Material parameters and classification of coupling effects}
\label{sec21}

The electromagnetic response in linear and time-invariant bianisotropic media at angular frequency $\omega$ is governed by the most general dyadic relations that are given in the frequency domain by 
\begin{equation}
\_D(\omega)=\overline{\overline{\epsilon}}_{\/F}(\omega)\cdot\_E(\omega)+\overline{\overline{\xi}}_{\/F}(\omega)\cdot\_H(\omega),\qquad
\_B(\omega)=\overline{\overline{\mu}}_{\/F}(\omega)\cdot\_H(\omega)+\overline{\overline{\zeta}}_{\/F}(\omega)\cdot\_E(\omega),
\label{bian_mr}
\end{equation} 
in which $\_D$ and $\_B$ are the electric and magnetic flux densities, $\_E$ and $\_H$ denote the electric and magnetic fields, and $\=\epsilon_{\/F}$ and $\=\mu_{\/F}$ represent the permittivity and permeability dyadics, respectively (the capital letter ``F" as a subscript indicates that these are the frequency-domain parameters). Magnetoelectric phenomena that can be caused by spatial dispersion or nonreciprocal effects are characterized by the coupling parameters $\overline{\overline{\xi}}_{\/F}$ and $\overline{\overline{\zeta}}_{\/F}$. For reciprocal media, these two parameters are related as $\overline{\overline{\xi}}_{\/F}=-{\overline{\overline{\zeta}}}_{\/F}^T$ ($T$ denotes the transpose operation)~\cite{Kong_book,Serdyukov}. Thus, it is convenient to write the coupling coefficients in the form 
\begin{equation}
\=\xi_{\/F}=\frac{1}{c}\Big(\=\chi_{\/F}-j\=\kappa_{\/F}\Big),
\qquad
\=\zeta_{\/F}=\frac{1}{c}\Big(\=\chi_{\/F}^T+j\=\kappa_{\/F}^T\Big).
\label{bian_rn}
\end{equation} 
Here, the dyadic $\=\kappa_{\/F}$ models reciprocal magnetoelectric effects, while the dyadic $\=\chi_{\/F}$ describes nonreciprocal effects. The factor $c$ is the speed of light, which is introduced to make the frequency-domain material parameters dimensionless. Moreover, the imaginary unit $``j"$ has been instigated to make all the material parameters real-valued for lossless media. Next, each of these parameters ($\=\kappa_{\/F}$ or $\=\chi_{\/F}$) is presented in the most general form as a summation of an isotropic and a trace-free dyadic. The reciprocal coupling parameter $\=\kappa_{\/F}$ is written as~\cite{JEWA_interactions,Materiatronics} 
\begin{equation} 
\=\kappa_{\/F}=\kappa_{\/F}\=I+\=M_{\/F}, 
\label{eq4} 
\end{equation}
where  $\kappa_{\/F}$ is $\frac{1}{3}$ of the trace of $\=\kappa_{\/F}$ (i.e.,~$\kappa_{\/F}=\frac{1}{3}{\rm{tr}}[\=\kappa_{\/F}]$), and $\=I$ is the unit dyadic. The remaining trace-free dyadic $\=M_{\/F}$ is decomposed
into its symmetric and antisymmetric parts: $\=M_{\/F}=\=N_{\/F}+\=J_{\/F}$. The symmetric part $\=N_{\/F}$ can be diagonalized so that $\=N_{\/F}=\sum_{i=1}^3\kappa_{{\/F}i}\_a_i\_a_i$ in which $\sum_{i=1}^3\kappa_{{\/F}i}=0$, and the antisymmetric part $\=J_{\/F}$ can be expressed in terms of a vector product operation as $\=J_{\/F}=\Omega_{\/F}\_b\times\=I$, where  $\_b$ is a unit vector defining the asymmetry axis. Notice that the parameters $\kappa_{\/F}$, $\kappa_{{\/F}i}$, and $\Omega_{\/F}$ are complex-valued factors defining the weights of each dyadic in the linear combination,  $\_a_i$ are the unit vectors in the basis of the eigenvectors of $\=\kappa_{\/F}$ (here, for simplicity, we take them to be real-valued). 
Likewise, we decompose the nonreciprocal coupling dyadic and write that 
\begin{equation} 
\=\chi_{\/F}=\chi_{\/F}\=I+\=P_{\/F}, 
\label{eq_nr} 
\end{equation} 
in which $\=P_{\/F}=\=Q_{\/F}+\=S_{\/F}$, $\=Q_{\/F}=\sum_{i=1}^3\chi_{{\/F}i}\_a_i\_a_i$, and, finally, $\=S_{\/F}=V_{\/F}\_b\times\=I$.
Of course, the unit vectors $\_a_i $ and $\_b$ are in general different from those associated with Eq.~\eqref{eq4}. 

The first term in Eq.~\eqref{eq4} defines isotropic \textit{true} chiral response \cite{condon1937theories}. It is nonzero only for three-dimensional molecules or meta-atoms with broken mirror symmetry. The three parameters $\kappa_{{\/F}i}$ quantify pseudo-chiral effects, such as optical activity in nonchiral samples at specific illumination directions \cite{sochava_chiral_1997,plum_extrinsic_2009}. The parameter $\Omega_{\/F}$ is called the omega-coupling coefficient. For nonreciprocal media, the first term in Eq.~\eqref{eq_nr} is the Tellegen parameter that models the isotropic nonreciprocal magnetoelectric effect (the Tellegen effect \cite{tellegen1948gyrator}). The parameters $\chi_{{\/F}i}$ are representing the pseudo-Tellegen property, and, eventually, the parameter $V_{\/F}$ is called the artificial velocity (see detailed discussions on bianisotropic material classification in Refs.~\cite{JEWA_interactions,Serdyukov,Materiatronics,Nanophotonics_review}). 

Thus, there are four classes of magnetoelectric coupling effects: chirality, omega coupling (both reciprocal), Tellegen, and artificial velocity (both nonreciprocal). Next, to set the ground for discussion of temporal inhomogeneities of coupling parameters, we will relate all of them with the distinct field effects at \emph{spatial} inhomogeneities. 

%%%%%%%%%%%%%%%%%%%%%%%%%%%%%%%%%%%%%%%%%%%%%%%%%%

\subsection{Fundamental field effects at bianisotropic spatial interfaces}
\label{ss_interfaces}

Let us consider spatial interfaces between half-spaces filled by lossless isotropic magnetodielectrics such as free space and bianisotropic media of different classes. For simplicity, we assume uniaxial structures, where the unit vector $\_b$ is orthogonal to the interface, and all the pseudo-chirality and pseudo-Tellegen parameters $\kappa_{{\/F}i}$ and $\chi_{{\/F}i}$ equal zero. The permittivity and permeability response is isotropic. Also, we suppose that a linearly polarized plane wave is normally incident from a bianisotropic medium on the interface. Under these conditions, the structure possesses uniaxial symmetry with the only preferred direction that is normal to the interface. Due to the magnetoelectric coupling in the medium in front of the interface, the following effects can take place (schematically illustrated in Fig.~\ref{fig_space} and with mathematical details given in the Appendix): 

\begin{figure*}[t!]
\centerline
{\includegraphics[width=0.9\linewidth]{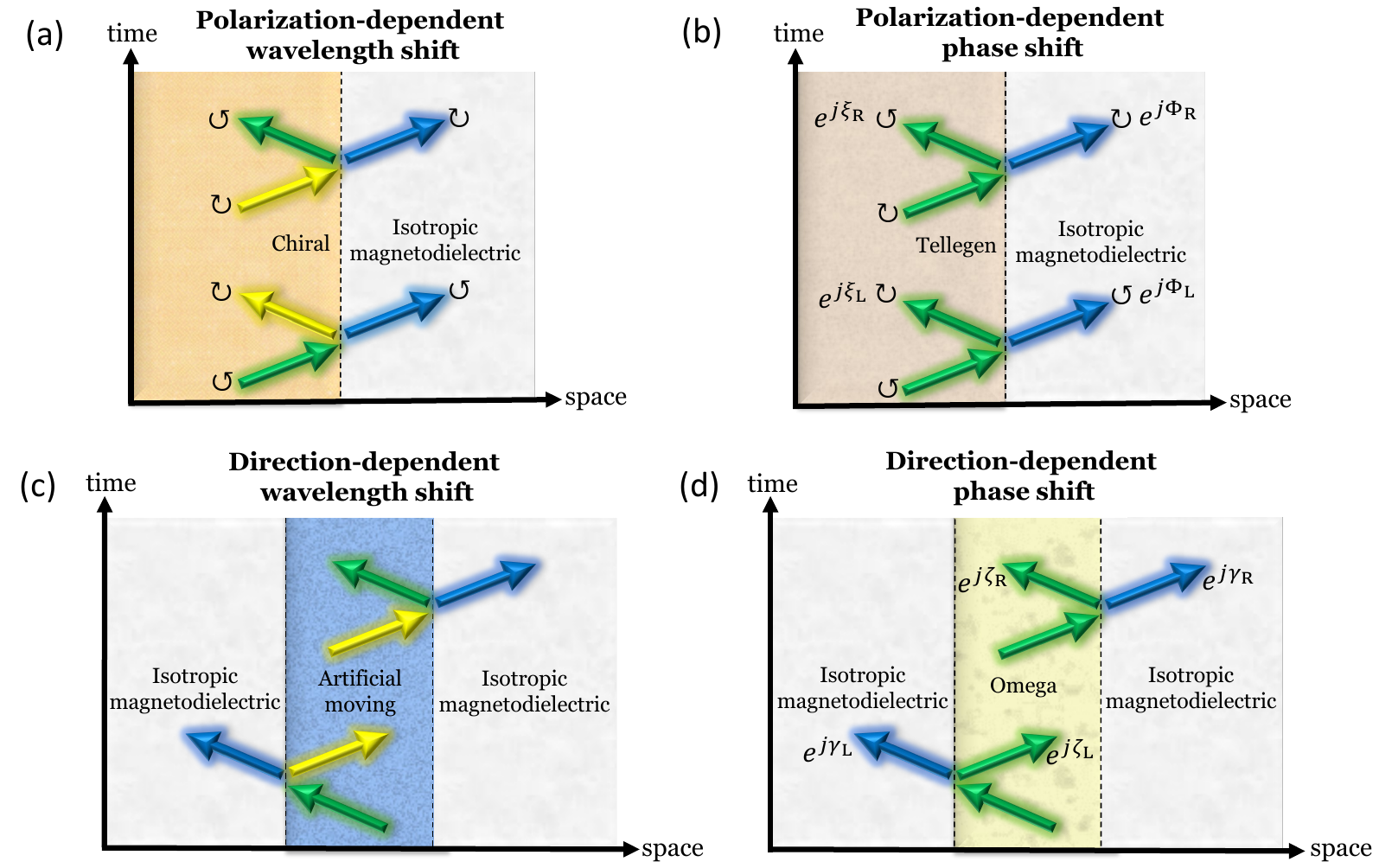}}
\caption{A schematic illustration of all scattering phenomena at single spatial interfaces between  bianisotropic media of different classes and an isotropic  magnetodielectric. Different colors indicate different wavelengths, rotating arrows indicate handedness of circular polarization, and the complex exponential indicates phase shifts taking place at the interface.} 
\label{fig_space}
\end{figure*} 

\begin{enumerate}

\item {Chiral coupling: If the medium is chiral, the polarization plane of linearly polarized incident waves continuously rotates around the propagation direction as the wave propagates within the medium (the optical activity effect). The wave transmitted into the isotropic magnetodielectric is also linearly polarized, but the polarization plane rotation stops. This effect is due to the fact that the right and left circularly polarized (RCP and LCP) components of both the incident and reflected waves have different propagation factors in the chiral media but the same propagation factor in the non-chiral magnetodielectric behind the interface.}

\item {Tellegen coupling: If the Tellegen parameter $\chi_{\/F}$ is real-valued (recall that we consider lossless media), linearly polarized incident waves remain linearly polarized in reflection and transmission, but their polarization directions abruptly rotate in the plane normal to the propagation direction. This effect takes place because the RCP and LCP components of the incident linearly polarized wave experience different phase shifts at the interface. Note that rotation of polarization in reflection is possible due to the nonreciprocal nature of the Tellegen effect.}

\item {Artificial velocity coupling: Here, the reflection and transmission coefficients are not dependent on the artificial velocity parameter, and there is no polarization transformation in reflection and transmission. However, waves traveling in artificially moving media in different directions have different phase velocities (and different decay factors if the medium is lossy). This is due to the fact that the propagation constant in such media depends on the artificial velocity parameter that controls the nonreciprocity of asymmetric coupling effects.}

\item {Omega coupling: Considering this class of magnetoelectric coupling, there is no polarization conversion, but the transmission and reflection coefficients for linearly polarized waves depend on the coupling parameter $\Omega_{\/F}$ so that if the sign changes ($\Omega_{\/F}\rightarrow-\Omega_{\/F}$), the corresponding phases of those coefficients also change sign. This means that the phase changes at the interface depend on the propagation direction of the incident wave. This effect is due to the fact that the characteristic impedance of plane waves in omega media depends on the propagation direction.} 

\end{enumerate}

We see that in this scenario, there are {\bf four} fundamental field effects 
measured by the co- and cross-polarized reflection and transmission coefficients and, interestingly, each of them is controlled by one of the {\bf four} parameters of the magnetoelectric material properties. 
%These basic phenomena are schematically illustrated in Fig.~\ref{fig_space}, and derivations of the corresponding reflection and transmission coefficients can be found in Appendix.

%%%%%%%%%%%%%%%%%%%%%%%%%%%%%%%%%%%%%%%%%%%%%%%%%%%%%%%%%%%%%%%%%%%%%%%%%%%%%%%%%%%%%%%%%%%%%%%%%%%%%%%

\subsection{Fundamental field effects at stationary bianisotropic metasurfaces}
\label{ss_ms}

Because some of the field effects are seen only in propagation over a finite distance (such as optical activity and nonreciprocal field decay), it is illustrative to list the main bianisotropy effects for thin bianisotropic layers or single arrays of densely packed bianisotropic particles. For this purpose, we use the known analytical solution for reflection and transmission coefficients for a dense array of uniaxial bianisotropic particles (period smaller than the wavelength) illuminated by normally incident linearly polarized waves~\cite{One-way}. The axis of the particles is normal to the array plane so that the whole structure has uniaxial symmetry. The solution for co- and cross-polarized reflected and transmitted waves reads  \cite{One-way}
\begin{equation} 
\begin{split}
&\mathbf{E}^{r}_{\rm co}=-\frac{j\omega}{2S}\left(\eta_0 {\alpha}_{\rm ee}^{\rm co}\pm 2j \Omega_{\/F} -\frac{1}{\eta_0}  {\alpha}_{\rm mm}^{\rm co}\right)\mathbf{E}_{\rm inc},\cr
&\mathbf{E}^{r}_{\rm cross}=-\frac{j\omega}{2S}\left(\eta_0 {\alpha}_{\rm ee}^{\rm cr}\mp 2 \chi_{\/F} -\frac{1}{\eta_0}  {\alpha}_{\rm mm}^{\rm cr}\right)\_a_z\times\mathbf{E}_{\rm inc},\cr
&\mathbf{E}^{t}_{\rm co}=
\left[1-\frac{j\omega}{2S}\left(\eta_0 {\alpha}_{\rm ee}^{\rm co}\pm 2  V_{\/F}
+\frac{1}{\eta_0} {\alpha}_{\rm mm}^{\rm co}\right)\right]
\mathbf{E}_{\rm inc},\cr
&\mathbf{E}^{t}_{\rm cross}=-\frac{j\omega}{2S} \left(\eta_0 {\alpha}_{\rm ee}^{\rm cr}\mp 2j  \kappa_{\/F}+\frac{1}{\eta_0}  {\alpha}_{\rm mm}^{\rm cr}\right) \_a_z\times\mathbf{E}_{\rm inc}.
\label{eq5}
\end{split}
\end{equation}
Here, $\_a_z$ is the unit vector normal to the array (we use a Cartesian coordinate system where the $z$-axis is along the propagation direction of the incident wave $\_E_{\rm{inc}}$), $ {\alpha}_{\rm ee,mm}^{\rm co, cr}$ are the diagonal and off-diagonal elements of the electric and magnetic effective polarizability dyadics. For clarity, we use the same notations for the coupling coefficients $ \kappa_{\/F} $, $ \Omega_{\/F} $, $ \chi_{\/F} $, and $  V_{\/F}$, although in Eqs.~\eqref{eq5} they have the meaning of effective (collective) polarizabilities of meta-atoms in 2D lattices~\cite{Karilainen,One-way}, in contrast to material parameters of volumetric, 3D media in Section~\ref{sec21}. 
%are the chirality, omega coupling, Tellegen coupling, and artificial velocity parameters, respectively. As before, all the parameters are expressed in the frequency domain (we drop the index ``F'' for compactness). 
We employ the electrical engineering convention $\exp(j\omega t)$ for describing harmonic time variations (notice that the same convention will be used in the rest of this paper). 
Furthermore, $\eta_0$ is the free-space intrinsic impedance, and $S$ is the unit-cell area.
The $\pm $ signs correspond to incident illuminations of the array from its opposite sides (that is, for the two opposite directions of incident wave propagation). 

Importantly, we see that if there is no magnetoelectric coupling ($ \kappa_{\/F}= \Omega_{\/F}= \chi_{\/F}=   V_{\/F}=0$), the response of the array is identical for both illumination directions. The presence of magnetoelectric coupling breaks this symmetry. Specifically, omega coupling makes the co-polarized reflection asymmetric, Tellegen coupling breaks the symmetry of cross-polarized reflection, artificial velocity makes the co-polarized transmission asymmetric, and, finally, chirality controls cross-polarized transmission (note that the unit vector $\_a_z$ is fixed for both illumination directions). 

We conclude that, complementing the properties of plane-wave reflection and transmission through a single interface, for uniaxial bianisotropic metasurfaces (with the axis normal to the metasurface plane and at normal incidence), there is a very similar correspondence between the field effects at such spatial inhomogeneities and the four magnetoelectric coupling parameters. Namely, there are two polarization-dependent effects: the chirality parameter controls polarization rotation in transmission, and the Tellegen parameter defines polarization rotation in reflection. In addition, there exist two direction-dependent phenomena where the polarization is conserved: omega coupling leads to asymmetry of reflection for illuminations of the metasurface from the opposite sides, while the artificial velocity parameter controls the corresponding asymmetry of the transmission of waves traveling in the opposite directions. In the last two cases, the reversal of the incident wave propagation direction is equivalent to the reversal of the coupling coefficient sign for the same incidence direction (see the list of phenomena at a single interface in subsection~\ref{ss_interfaces}). These four fundamental effects and their correspondence to the four classes of bianisotropic coupling are discussed in detail in Ref.~\cite{Nanophotonics_review} (an illustration is provided by Fig.~12 in that review paper). 

In the following, our goal is to study the effects of \emph{temporal} interfaces in bianisotropic media, where the magnetoelectric coupling parameters rapidly change in time, instead of exhibiting sharp inhomogeneities in space.

%%%%%%%%%%%%%%%%%%%%%%%%%%%%%%%%%%%%%%%%%%%%%%%%%%%%%%%%%%%%%% Section 3  %%%%%%%%%%%%%%%%%%%%%%%%%%%%%%%%%%%%%%%%%%%%%%%%%%%%%%%%%%%%%%
%%%%%%%%%%%%%%%%%%%%%%%%%%%%%%%%%%%%%%%%%%%%%%%%%%%%%%%%%%%%%%

\section{Temporal Interfaces} 
\label{sec:TIBMA4}

In this section, we focus on a temporal interface between the bianisotropic media of the four fundamental classes and an isotropic magnetodielectric medium. 
To study the related field phenomena, first, we will need material relations written in the time domain, because the boundary conditions at time interfaces are defined for field values at a certain moment in time. To this end, we will use the Condon-Tellegen form of material relations.

\subsection{Condon-Tellegen material relations}

The time-domain counterparts of Eq.~\eqref{bian_mr}, even for time-invariant media, contain convolution integrals due to nonlocality in time dubbed as temporal or frequency dispersion~\cite{jackson1999classical}. Only if we neglect such frequency dispersion, the relations become simple enough and allow analytical solutions for identifying and classifying possible field effects. For nonreciprocal magnetoelectric coupling material parameters (the Tellegen parameter and the artificial velocity), we can assume that their frequency dispersion is negligible and use the material relations in the Tellegen form also in the time domain (see Eqs.~\eqref{bian_mr} and \eqref{bian_rn}). However, this is not possible for the reciprocal coupling coefficients, because the very nature of these effects is spatial dispersion in the medium. Therefore, in the frequency domain, these parameters must depend on the frequency. In particular, in the limit of zero frequency, they always tend to zero as linear functions of the frequency~\cite{Serdyukov}. For this reason, we employ the Condon  model~\cite{condon1937theories,sihvola1991bi} to describe the corresponding coupling effects in the time domain and write the general bianisotropic time-domain material relations in the form 
\begin{equation} 
\begin{split}
&\_D(t)=\overline{\overline{\epsilon}}\cdot\_E(t)+\frac{\chi}{c}\,\_H(t)-\frac{\kappa}{c}\,\frac{\partial\_H(t)}{\partial t}+\frac{V}{c}\mathbf{a}_z\times\mathbf{H}(t)+\frac{\Omega}{ c}\_a_z\times\frac{\partial\_H(t)}{\partial t},\cr
&\_B(t)=\overline{\overline{\mu}}\cdot\_H(t)+\frac{\chi}{c}\,\_E(t)+\frac{\kappa}{c}\,\frac{\partial\_E(t)}{\partial t}-\frac{V}{c}\mathbf{a}_z\times\mathbf{E}(t)+\frac{\Omega}{c}\_a_z\times\frac{\partial\_E(t)}{\partial t}. 
\end{split}
\label{eq:CTMS} 
\end{equation}
It is worth noting that since these expressions are applicable at any point in space, for brevity, we did not include the position vector as an independent variable for the fields ($\_E,\,\_H$) and the flux densities ($\_D,\,\_B$). In Eq.~\eqref{eq:CTMS}, we observe that while the nonreciprocal terms $\chi$ and $V$ are dimensionless (similarly to their frequency-domain counterparts $\chi_{\/F}$ and $V_{\/F}$ described in Section~\ref{sec:BMICLTI}), the dimension of the reciprocal terms $\kappa$ and $\Omega$ is now second. In this model, the assumption is that the field oscillations are at frequencies that are well below the resonances of material response. This is a physically valid model that properly accounts for the inevitable frequency dispersion of chirality and omega coupling. In particular, for temporally constant fields ($\partial\_E/\partial t=\partial\_H/\partial t=0$), the reciprocal coupling vanishes, which corresponds to the fact that there is neither chiral nor omega coupling in statics. 

%%%%%%%%%%%%%%%%%%%%%%%%%%%%%%%%%%%%%%%%%%%%%%%%%%%%%%

\subsection{Polarization-dependent phenomena}  

We start from studying effects at rapid changes of the magnetoelectric coefficients that define isotropic coupling: chirality and Tellegen coupling. At spatial interfaces, these parameters control polarization-dependent phenomena (see Sections~\ref{ss_interfaces} and \ref{ss_ms}), and it is expected that novel polarization-dependent scattering effects can exist at temporal interfaces. 

\subsubsection{Chiral temporal interfaces}

Field effects at rapid changes of chirality parameters in isotropic chiral media were already considered in Ref.~\cite{mostafa2023spin}, and possible resonant dispersion effects were discussed in Ref.~\cite{Alu_copy}. Here, we present the main results for completeness of the study, applying the time-domain Condon model. However, we stress that different from Ref.~\cite{mostafa2023spin}, we consider a time discontinuity between chiral medium and any magnetodielectric one that has arbitrary values of permittivity and permeability. It is shown that this difference, for example, results in having nonzero reflected waves. To demonstrate how one can manipulate the polarization states by using only chiral temporal interfaces, Eq.~\eqref{eq:CTMS} is rewritten as
\begin{equation}
\_D=\epsilon\_E+\frac{\kappa}{c}\frac{\partial\_H}{\partial t},\quad
\_B=\mu\_H-\frac{\kappa}{c}\frac{\partial\_E}{\partial t},
\end{equation}
where $\kappa$ is the chirality parameter. We assume that a linearly polarized plane wave is propagating in the chiral medium. As is well known, the phase constants of right-handed and left-handed circularly polarized (RCP and LCP) wave components are different: $\beta_{\/R}=\omega(\sqrt{\mu\epsilon}-\omega\kappa/c)$ and $\beta_{\/L}=\omega(\sqrt{\mu\epsilon}+\omega\kappa/c)$. Presenting a linearly polarized wave as a sum of RCP and LCP components, we write the electric and magnetic fields as 
\begin{equation}
\_E={E_0\over2}\Big[(\_a_x-j\_a_y)\exp(-j\beta_{\/R}z)+(\_a_x+j\_a_y)\exp(-j\beta_{\/L}z)\Big]\exp(j\omega t)
\end{equation}
and  
\begin{equation} 
\_H={jE_0\over2\mu}\Bigg[\Big({\beta_{\/R}\over\omega}+{\omega\kappa\over c}\Big)(\_a_x-j\_a_y)\exp(-j\beta_{\/R}z)-\Big({\beta_{\/L}\over\omega}-{\omega\kappa\over c}\Big)(\_a_x+j\_a_y)\exp(-j\beta_{\/L}z)\Bigg]\exp(j\omega t).
\label{eq:HMAGCHI}
\end{equation}

In the following, we study a temporal interface at $t=0$ at which the chiral medium is replaced by an isotropic magnetodielectric medium described effectively by permittivity $\epsilon_{\/{MD}}$ and permeability $\mu_{\/{MD}}$. 
After some transition period, the fields in this simple magnetodielectric will be in the form of its eigenwaves. Thus, the resulting flux densities for both RCP and LCP components are expressed as 
\begin{equation}
\begin{split}
&\_D_{\/{MD}}=\epsilon_{\/{MD}}\Big(\_E^r\exp(-j\omega_{\/{MD}}t)+\_E^t\exp(j\omega_{\/{MD}}t)\Big)\exp(-j\beta z),\cr
&\_B_{\/{MD}}=\sqrt{\mu_{\/{MD}}\epsilon_{\/{MD}}}\_a_z\times\Big(\_E^t\exp(j\omega_{\/{MD}}t)-\_E^r\exp(-j\omega_{\/{MD}}t)\Big)\exp(-j\beta z),
\end{split}
\label{eq:DBMDIMV}
\end{equation} 
where the propagation constant $\beta$ is conserved at time interfaces in spatially uniform media. Equation~\eqref{eq:DBMDIMV} indicates that due to the presence of the temporal interface, there are reflected (or backward) $\_E^r$ and transmitted (or forward) $\_E^t$ waves that simultaneously propagate in the magnetodielectric medium. 

Next, we calculate the flux densities and determine the amplitudes of the forward and backward waves by imposing the time-jump boundary conditions (continuity of $\_D$ and $\_B$). However, before doing that, it is important to note that the conservation of the phase constant and that $\beta_{\/R}\neq\beta_{\/L}$ lead to the fact that the two corresponding sets of forward and backward waves will propagate at two distinct angular frequencies. Indeed, for the left-handed polarization, we have $\omega_{\/{MDL}}=(\omega/\sqrt{\mu_{\/{MD}}\epsilon_{\/{MD}}})(\sqrt{\mu\epsilon}+\omega\kappa/c)$, and for the right-handed polarization, we obtain $\omega_{\/{MDR}}=(\omega/\sqrt{\mu_{\/{MD}}\epsilon_{\/{MD}}})(\sqrt{\mu\epsilon}-\omega\kappa/c)$. As a consequence, after the temporal jump, the total electric field should be written as $\_E=\big[\_E^r_{\/L}\exp(-j\omega_{\/{MDL}}t)+\_E^t_{\/L}\exp(j\omega_{\/{MDL}}t)\big]\exp(-j\beta_{\/L}z)+\big[\_E^r_{\/R}\exp(-j\omega_{\/{MDR}}t)+\_E^t_{\/R}\exp(j\omega_{\/{MDR}}t)\big]\exp(-j\beta_{\/R}z)$. Knowing 
the total electric field, we readily revise the flux densities in Eq.~\eqref{eq:DBMDIMV}. Omitting intermediate derivations, the final results take the form  
\begin{equation}
\_E^r_{\/R}={E_0\over4}M^r\Big(1-{\omega\kappa\over c\sqrt{\mu\epsilon}}\Big)(\_a_x-j\_a_y),\quad
\_E^r_{\/L}={E_0\over4}M^r\Big(1+{\omega\kappa\over c\sqrt{\mu\epsilon}}\Big)(\_a_x+j\_a_y)
\end{equation}
and 
\begin{equation}
\_E^t_{\/R}={E_0\over4}M^t\Big(1-{\omega\kappa\over c\sqrt{\mu\epsilon}}\Big)(\_a_x-j\_a_y),\quad
\_E^t_{\/L}={E_0\over4}M^t\Big(1+{\omega\kappa\over c\sqrt{\mu\epsilon}}\Big)(\_a_x+j\_a_y),
\end{equation}
in which $M^r=[(\epsilon/\epsilon_{\/{MD}})-(\sqrt{\mu\epsilon}/\sqrt{\mu_{\/{MD}}\epsilon_{\/{MD}}})]$ and $M^t=[(\epsilon/\epsilon_{\/{MD}})+(\sqrt{\mu\epsilon}/\sqrt{\mu_{\/{MD}}\epsilon_{\/{MD}}})]$ (notice that the parameter $M$ is the same as calculated by F.~R.~Morgenthaler when he studied temporal discontinuities between two magnetodielectric media~\cite{morgenthaler1958velocity}). 

In summary, the upshot of such a chiral temporal interface is the splitting of a linearly polarized incident wave into two circularly polarized waves with opposite handedness that propagate at the same phase velocity ($v_{\/{MD}}=1/\sqrt{\mu_{\/{MD}}\epsilon_{\/{MD}}})$ but at different angular frequencies. The frequency difference is  equal to $\Delta=\omega_{\/{MDL}}-\omega_{\/{MDR}}=2\omega^2\kappa/(c\sqrt{\mu_{\/{MD}}\epsilon_{\/{MD}}})$. It is clear that a higher refractive index of the magnetodielectric medium results in a smaller value for $\Delta$. On the other hand, we note interesting extreme translations into magnetodielectric media with near-zero values of permeability or permittivity. 

We would like to emphasize that the above conclusion reminds the renowned work done by Otto Stern and Walther Gerlach almost one hundred years ago. While the above theory shows that electromagnetic waves of orthogonal circular polarizations (of opposite ``spins'') are converted into different frequencies at time interfaces, in the Stern-Gerlach experiment, a beam of neutral atoms carrying nonzero magnetic moments is split at a spatial discontinuity. The spin-up and spin-down components of the beam after the discontinuity correspond to two different momenta, which gives rise to the propagation of these two components in two different directions in space~\cite{GriffithsQM}.

\subsubsection{Tellegen temporal interfaces}

Let us next consider a Tellegen medium whose constitutive relations are given by Eq.~\eqref{eq:CTMS} as  
\begin{equation}
\_D=\epsilon\_E+\frac{\chi}{c}\_H,\quad\_B=\mu\_H+\frac{\chi}{c}\_E,
\label{eq:TDB}
\end{equation}
in which $\epsilon$ and $\mu$ are the effective permittivity and permeability of the medium, respectively, and $\chi$ is the Tellegen parameter. We again assume that a linearly polarized plane wave is propagating within the medium in the $z$-direction, and its electric field is written as $\_E=E_0\exp(-j\beta z)\exp(j\omega t)\_a_x$. Here, $E_0$ is the field amplitude, $\beta$ represents the phase constant, and $\omega$ denotes the angular frequency. This plane wave can be viewed as a combination of right-handed and left-handed circularly polarized plane waves whose electric fields are expressed as 
\begin{equation}
\_E_{\/R}={E_0\over2}(\_a_x-j\_a_y)e^{-j\beta z}e^{j\omega t},\quad
\_E_{\/L}={E_0\over2}(\_a_x+j\_a_y)e^{-j\beta z}e^{j\omega t}.
\label{ET:eq}
\end{equation}
Note that, unlike in the context of a chiral medium, the phase constants of the RCP and LCP components are equal for a Tellegen medium. To calculate this phase constant $\beta$ and the corresponding magnetic fields $\_H_{\/R,\/L}$, we need to use Maxwell's equations. By doing that, we find 
\begin{equation}
\_H_{\/R}={E_0\over2\mu}\Big({j\beta\over\omega}-{\chi\over c}\Big)(\_a_x-j\_a_y)e^{-j\beta z}e^{j\omega t},\quad 
\_H_{\/L}=-{E_0\over2\mu}\Big({j\beta\over\omega}+{\chi\over c}\Big)(\_a_x+j\_a_y)e^{-j\beta z}e^{j\omega t},
\label{HT:eq}
\end{equation}
where $\beta=\omega\sqrt{\mu\epsilon-\chi^2/c^2}$. If we contemplate Eqs.~\eqref{ET:eq} and \eqref{HT:eq}, we observe that, importantly, the total incident magnetic field is not perpendicular to the total incident electric field. While the electric field has only one component in the $x$ direction, the magnetic field possesses two components in the $xy$ plane. Later, we see the effect of this salient feature on the fields after the Tellegen parameter vanishes abruptly in time. Since we know  the total electric and magnetic fields, Eq.~\eqref{eq:TDB} provides us with the flux densities which are readily simplified as 
\begin{equation}
\_D=\frac{E_0}{2}\Bigg[\epsilon-{\chi^2\over\mu c^2}\pm j{\chi\over\mu c}\sqrt{\mu\epsilon-{\chi^2\over c^2}}\Bigg](\_a_x\mp j\_a_y)e^{-j\beta z}e^{j\omega t},\quad 
\_B=\pm j\frac{E_0}{2}\sqrt{\mu\epsilon-{\chi^2\over c^2}}(\_a_x\mp j\_a_y)e^{-j\beta z}e^{j\omega t}.
\end{equation}
Here, the upper and lower signs correspond to the right-handed (R) and left-handed (L) circularly polarized components, respectively. 

Looking for the fields after the time jump to an isotropic magnetodielectric in the form illustrated by Eq.~\eqref{eq:DBMDIMV}, we use the dispersion relations to find that the resulting forward and backward waves have a different angular frequency as compared to the angular frequency of the incident wave. Indeed, $\omega\sqrt{\mu\epsilon-\chi^2/c^2}=\omega_{\/{MD}}\sqrt{\mu_{\/{MD}}\epsilon_{\/{MD}}}$ which results in $\omega_{\/{MD}}=\omega\sqrt{(\mu\epsilon-\chi^2/c^2)/(\mu_{\/{MD}}\epsilon_{\/{MD}})}$. We see that if the refractive index of the magnetodielectric medium is large, satisfying $n_{\/{MD}}>\sqrt{c^2\mu\epsilon-\chi^2}$, the angular frequency is red shifted. Otherwise, if $n_{\/{MD}}<\sqrt{c^2\mu\epsilon-\chi^2}$, we have a frequency conversion to a higher value. Also, notice that for the special value of the Tellegen parameter $\chi=c\sqrt{\mu\epsilon}$, the resulting field distribution is static in time ($\omega_{\/{MD}}=0$). To find the amplitudes of the forward and backward waves, we apply the boundary conditions that are the continuity of the electric and magnetic flux densities at time $t=0$ for any point $z$ in space: $\_D(z,0)=\_D_{\/{MD}}(z,0)$ and $\_B(z,0)=\_B_{\/{MD}}(z,0)$. After some algebraic manipulations, we deduce that 
\begin{equation}
\begin{split}
&\_E^r_{{\/R},{\/L}}={E_0\over4}\Bigg({\epsilon\over\epsilon_{\/{MD}}}-\sqrt{{\mu\epsilon\over\mu_{\/{MD}}\epsilon_{\/{MD}}}-{\chi^2\over\mu_{\/{MD}}\epsilon_{\/{MD}}c^2}}-{\chi^2\over\mu\epsilon_{\/{MD}}c^2}\pm j{\chi\over\mu\epsilon_{\/{MD}}c}\sqrt{\mu\epsilon-{\chi^2\over c^2}}\Bigg)(\_a_x\mp j\_a_y),\cr 
&\_E^t_{{\/R},{\/L}}={E_0\over4}\Bigg({\epsilon\over\epsilon_{\/{MD}}}+ \sqrt{{\mu\epsilon\over\mu_{\/{MD}}\epsilon_{\/{MD}}}-{\chi^2\over\mu_{\/{MD}}\epsilon_{\/{MD}}c^2}}-{\chi^2\over\mu\epsilon_{\/{MD}}c^2}\pm j{\chi\over\mu\epsilon_{\/{MD}}c}\sqrt{\mu\epsilon-{\chi^2\over c^2}}\Bigg)(\_a_x\mp j\_a_y).
\end{split}
\end{equation}
This equation demonstrates that the imaginary parts of the expressions inside the brackets are different for different circularly polarized components. In fact, the expressions associated with the right-handed component are complex conjugates of the ones related to the left-handed component. Now, the total fields after the temporal discontinuity are found by simply adding the fields of the two polarization states. Thus, ultimately, we have      
\begin{equation}
\begin{split}
&\_E^r={E_0\over2}\Bigg({\epsilon\over\epsilon_{\/{MD}}}-\sqrt{{\mu\epsilon\over\mu_{\/{MD}}\epsilon_{\/{MD}}}-{\chi^2\over\mu_{\/{MD}}\epsilon_{\/{MD}}c^2}}-{\chi^2\over\mu\epsilon_{\/{MD}}c^2}\Bigg)\_a_x+{E_0\over2}{\chi\over\mu\epsilon_{\/{MD}}c}\sqrt{\mu\epsilon-{\chi^2\over c^2}}\_a_y,\cr 
&\_E^t={E_0\over2}\Bigg({\epsilon\over\epsilon_{\/{MD}}}+\sqrt{{\mu\epsilon\over\mu_{\/{MD}}\epsilon_{\/{MD}}}-{\chi^2\over\mu_{\/{MD}}\epsilon_{\/{MD}}c^2}}-{\chi^2\over\mu\epsilon_{\/{MD}}c^2}\Bigg)\_a_x+{E_0\over2}{\chi\over\mu\epsilon_{\/{MD}}c}\sqrt{\mu\epsilon-{\chi^2\over c^2}}\_a_y.
\end{split}
\label{eq:ETRLPR}
\end{equation}
This result explicitly shows that after making the temporal jump at $t=0$, the electric field rotates instantaneously in the $xy$ plane, although the corresponding plane wave is still linearly polarized. The sense of rotation is defined by the sign of the Tellegen parameter $\chi$. Quite interestingly, according to Eq.~\eqref{eq:ETRLPR}, it is possible to have a $90^\circ$ rotation for the electric field $\_E^r$. For that, the $x$ component must disappear, which gives rise to the following condition: $\chi^2=c^2(\mu\epsilon-\mu^2\epsilon_{\/{MD}}/\mu_{\/{MD}})>0$ in which $(\mu/\epsilon)<(\mu_{\/{MD}}/\epsilon_{\/{MD}})$. Under this condition, the created backward wave has only a $y$ component. 

%%%%%%%%%%%%%%%%%%%%%%%%%%%%%%%%%%%%%%%%%%%%%%%

%%%%%%%%%%%%%%%%%%%%%%%%%%%%%%%%%%%%%%%%%%%%
%%%%%%%%%%%%%%%%%%%%%%%%%%%%%%%%%%%%%%%%%%%%

\subsection{Direction-dependent phenomena} 
\label{subsec:DDPMO}

Next, we scrutinize directional-dependent phenomena, which means that scattering of waves at a temporal interface depends on the propagation direction of the incident wave in the bianisotropic medium. 
 
\subsubsection{Temporal interfaces in artificial moving media}
\label{subsTIAMM}

Let us first consider time jumps in nondispersive artificial moving media. According to the Tellegen model of  Eq.~\eqref{eq:CTMS}, the constitutive relations read 
\begin{equation}
\mathbf{D}=\epsilon\mathbf{E}+\frac{V}{c}\mathbf{a}_z\times\mathbf{H},\quad\mathbf{B}=\mu\mathbf{H}-\frac{V}{c}\mathbf{a}_z\times\mathbf{E},
\end{equation} 
in which $\mathbf{a}_z$ is the unit vector along the $z$-axis, and $V$ represents the dimensionless coupling parameter called artificial velocity. Once again, we assume that a linearly polarized plane wave is propagating in this bianisotropic medium in the $+z$-direction: $\mathbf{E}=E_0\exp(-j\beta z)\exp(j\omega t)\mathbf{a}_x$. Subsequently, Maxwell's equations define the corresponding magnetic field as $\mathbf{H}=(1/\mu)((\beta/\omega)+(V/c))E_0\exp(-j\beta z)\exp(j\omega t)\mathbf{a}_y$ and the corresponding phase constant as $\beta=\omega(\sqrt{\mu\epsilon}-V/c)$. Finally, we derive the electric and magnetic flux densities, which are simplified to
\begin{equation}
\mathbf{D}=\Big(\epsilon-\frac{\sqrt{\epsilon}}{\sqrt{\mu}}{V\over c}\Big)E_0\exp(-j\beta z)\exp(j\omega t)\mathbf{a}_x,\quad 
\mathbf{B}=\Big(\sqrt{\mu\epsilon}-{V\over c}\Big)E_0\exp(-j\beta z)\exp(j\omega t)\mathbf{a}_y.
\end{equation} 
At $t=0$, we presume that a fast transformation of the artificial moving medium to an isotropic magnetodielectric one occurs (i.e., the coupling parameter $V$ becomes zero, and the effective permittivity and permeability in general also change). The first sequel to such transformation is a frequency conversion, which means that the scattered plane waves have the following angular frequency: $\omega_{\/{MD}}=(\omega/\sqrt{\mu_{\/{MD}}\epsilon_{\/{MD}}})(\sqrt{\mu\epsilon}-V/c)$. The second consequence is the simultaneous creation of forward and backward propagating waves. Remembering Eq.~\eqref{eq:DBMDIMV} and keeping in mind the boundary conditions, we deduce the amplitudes of the backward and forward waves:
\begin{equation}
\_E^r=\frac{1}{2}\Bigg[M^r-\frac{\sqrt{\epsilon}}{\sqrt{\mu}}\frac{V}{c\epsilon_{\/{MD}}}+\frac{V}{c\sqrt{\mu_{\/{MD}}\epsilon_{\/{MD}}}}\Bigg]\_a_x,\quad 
\_E^t=\frac{1}{2}\Bigg[M^t-\frac{\sqrt{\epsilon}}{\sqrt{\mu}}\frac{V}{c\epsilon_{\/{MD}}}-\frac{V}{c\sqrt{\mu_{\/{MD}}\epsilon_{\/{MD}}}}\Bigg]\_a_x.
\label{tau-gamma}
\end{equation} 
As a simple check, we explicitly see that if the coupling parameter $V$ is zero ($V=0$), we achieve exactly the same expressions as derived by Morgenthaler~\cite{morgenthaler1958velocity}. 

Let us next discuss what happens if a uniform plane wave is propagating in the opposite direction before the temporal jump (i.e., in the $-z$ direction). To do that, we do not need to re-calculate all the previous steps. What we need is to only reverse the sign of the artificial velocity parameter $V$: $V\rightarrow-V$. This substitution transforms the waves and the angular frequency after the jump to  
\begin{equation}
\_E^r_{\leftarrow}={1\over2}\Bigg[M^r+\frac{\sqrt{\epsilon}}{\sqrt{\mu}}\frac{V}{c\epsilon_{\/{MD}}}-\frac{V}{c\sqrt{\mu_{\/{MD}}\epsilon_{\/{MD}}}}\Bigg]\_a_x,\quad
\_E^t_{\leftarrow}={1\over2}\Bigg[M^t+\frac{\sqrt{\epsilon}}{\sqrt{\mu}}
\frac{V}{c\epsilon_{\/{MD}}}+\frac{V}{c\sqrt{\mu_{\/{MD}}\epsilon_{\/{MD}}}}\Bigg]\_a_x,
\end{equation}
and 
$\omega_{{\/{MD}}\leftarrow}=(\omega/\sqrt{\mu_{\/{MD}}\epsilon_{\/{MD}}})(\sqrt{\mu\epsilon}+V/c)$. If we compare these results with the expressions derived for the plane wave propagating in the $+z$ direction, we observe direction-dependent effects. Depending on the direction of propagation in the moving medium, the angular frequency shifts differently, and the amplitude of the forward or backward electric field increases or decreases with respect to the artificial velocity.

%%%%%%%%%%%%%%%%%%%%%%%%%%%%%%%%%%%%%%%%%%%%%%%%%%%%%%%%%%%%%%

%\noindent\textbf{\textcolor{blue}{Omega temporal interfaces.}} 

\subsubsection{Omega temporal interfaces}

Finally, we study the last scenario of omega temporal interfaces and demonstrate that such an interface also provides direction-dependent scattering. Considering a lossless uniaxial omega medium with an axis along $\_a_z$, the time-domain Condon constitutive relations are expressed as
\begin{equation} 
\_D=\epsilon\_E+{\Omega\over c}\_a_z\times\frac{\partial\_H}{\partial t},\quad
\_B=\mu\_H+{\Omega\over c}\_a_z\times\frac{\partial\_E}{\partial t},
\label{eq:CROC}
\end{equation}
in which $\Omega$ is the real-valued omega coupling coefficient. For transverse electromagnetic plane waves, having linear polarization and propagating along the medium axis $\_a_z$, we solve Maxwell's equations and derive the fields, which are written as $\mathbf{E}=E_0\exp(-j\beta z)\exp(j\omega t)\mathbf{a}_x$ and $\_H=(E_0/\mu)((\beta/\omega)-(j\omega\Omega/c))\exp(-j\beta z)\exp(j\omega t)\_a_y$, where $\beta=\omega\sqrt{\mu\epsilon-\omega^2\Omega^2/c^2}$. This expression manifests that the phase constant does not change if the sign of $\Omega$ is reversed. Thus, regardless of the propagation direction, $\beta$ is fixed, and, hence, the angular frequency conversion due to the temporal discontinuity does not depend on the propagation direction of the incident wave (in contrast to an artificial moving medium): $\omega_{{\/{MD}}\leftrightarrow}=(\omega/\sqrt{\mu_{\/{MD}}\epsilon_{\/{MD}}})\sqrt{\mu\epsilon-\omega^2\Omega^2/c^2}$. We can now deduce the electric and magnetic flux densities, and, similarly to the previous derivations, calculate the amplitudes of the scattered forward and backward waves. After some algebraic manipulations, we derive that 
\begin{equation}
\_E^r={1\over2}\Bigg[{\epsilon\over\epsilon_{\/{MD}}}-{\sqrt{\mu\epsilon-\omega^2(\Omega/c)^2}\over\sqrt{\mu_{\/{MD}}\epsilon_{\/{MD}}}}-{\omega^2(\Omega/c)^2\over\mu\epsilon_{\/{MD}}}-j{\omega(\Omega/c)\over\mu\epsilon_{\/{MD}}}\sqrt{\mu\epsilon-\omega^2(\Omega/c)^2}\Bigg]\_a_x
\end{equation}
and 
\begin{equation}
\_E^t={1\over2}\Bigg[{\epsilon\over\epsilon_{\/{MD}}}+{\sqrt{\mu\epsilon-\omega^2(\Omega/c)^2}\over\sqrt{\mu_{\/{MD}}\epsilon_{\/{MD}}}}-{\omega^2(\Omega/c)^2\over\mu\epsilon_{\/{MD}}}-j{\omega(\Omega/c)\over\mu\epsilon_{\/{MD}}}\sqrt{\mu\epsilon-\omega^2(\Omega/c)^2}\Bigg]\_a_x.
\end{equation} 
The above equations show that while the real parts of the $x$ component of the fields are even with respect to $\Omega$, alluringly, the imaginary parts are odd functions. Therefore,  if the incident plane wave propagates in the opposite direction (i.e., $-z$ direction), the corresponding phases of the electric fields change sign ($\angle\_E^r=-\angle\_E^r_{\leftarrow}$ and $\angle\_E^t=-\angle\_E^t_{\leftarrow}$). However, notice that the magnitudes of the resulting fields are the same.    

%%%%%%%%%%%%%%%%%%%%%%%%%%%%%%%%%%%%%%%%%%%%%%%%%%%%%%
%%%%%%%%%%%%%%%%%%%%%%%%%%%%%%%%%%%%%%%%%%%%%%%%%%%%%% 

\subsection{Field evaporation} 

The above theory provides general expressions for fields after temporal discontinuities of all possible coupling coefficients.  Here, we inspect one particular scenario that is related to the case of zero effective indices of bianisotropic media. Recall that at the angular frequency where the effective index of a medium becomes zero, the plane wave does not vary in space, while it oscillates in time (see, e.g., Ref.~\cite{liberal2017ZINDEX}).

For simplicity, let us concentrate only on the case of a temporal interface between an artificial moving medium and a magnetodielectric one, and assume that the artificial moving medium is characterized by the coupling parameter $V=c\sqrt{\mu\epsilon}$. This specific value corresponds to a zero phase constant and, accordingly, zero effective index for one of the wave propagation directions (see subsection~\ref{subsTIAMM}). Based on the equations that we derived for the fields, Eq.~\eqref{tau-gamma}, we explicitly observe that in this case, both the backward and forward waves vanish after the temporal jump. Therefore, after the jump, there is no propagating power or stored energy in the medium. Furthermore, at such temporal discontinuity, the frequency of the incident wave is converted to zero (i.e., $\omega_{\/{MD}}=0$). Notice that this phenomenon takes place for arbitrary values of the permittivity and permeability of the magnetodielectric medium. It appears that since in this case, the fields of the original plane wave completely disappear, its energy is fully transferred to the device that changes the material properties. 

We coin the above oddity as ``field evaporation". This effect happens also for temporal discontinuities of chirality, Tellegen, and omega coupling coefficients, in which cases the special values are equal to 
\begin{equation} 
\kappa=\frac{c}{\omega}\sqrt{\mu\epsilon}, \quad \chi=c\sqrt{\mu\epsilon}, \quad \Omega=\frac{c}{\omega}\sqrt{\mu\epsilon}.
\end{equation} 
For chiral media, however, field evaporation occurs only for one of the circularly polarized waves at the angular frequency $\omega$.
Likewise, evaporation takes place only for one of the two opposite propagation directions of the incident wave in artificial moving media. For the other propagation direction, at the particular value of $V/c=\sqrt{\mu\epsilon}$, the fields $\_E^r_{\leftarrow}$ and $\_E^t_{\leftarrow}$ do not vanish, and $\omega_{{\/{MD}}\leftarrow}=2\omega\sqrt{\mu\epsilon}/\sqrt{\mu_{\/{MD}}\epsilon_{\/{MD}}}$.
At these special values of the coupling parameters, the symmetry breaking at temporal interfaces is the strongest in the relative sense, as the incident waves of one of the two polarizations or propagation directions produce zero fields after the time discontinuity.

%%%%%%%%%%%%%%%%%%%%%%%%%%%%%%%%%%%%%%%%%%%%%%%%%%%%%%
%%%%%%%%%%%%%%%%%%%%%%%%%%%%%%%%%%%%%%%%%%%%%%%%%%%%%% 

\begin{figure}[t]
\centerline 
{\includegraphics[width=0.9\linewidth]{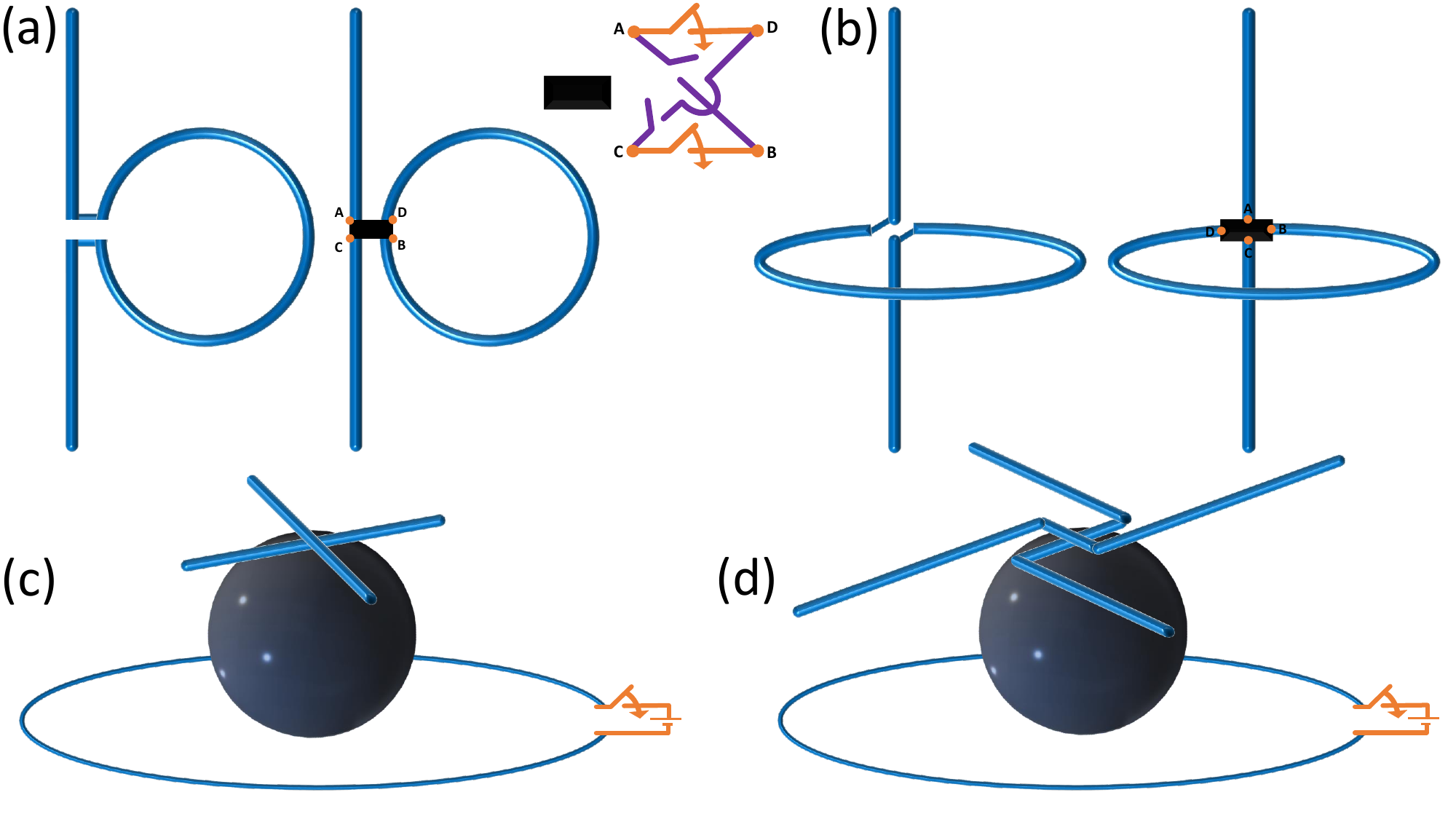}} 
\caption{Viable microwave-range realizations of time-varying bianisotropic particles. (a) and (b)--The sign of the omega or chirality parameter of canonical omega particles or spirals can be changed by using electronic switches. (c) and (d)--Tellegen and artificial velocity effects can be switched off by switching off the external magnetic bias field. Here, the circular wire shown by the loop with blue color carries static electric current.}
\label{fig_spirals}
\end{figure}

\subsection{About possible realizations of time-varying bianisotropic media} 

Here, we briefly discuss possible realizations of time-modulated magnetoelectric coupling. 
Figure~\ref{fig_spirals} illustrates some possibilities to create meta-atoms, metasurfaces, and metamaterials with electrically switchable or tunable magnetoelectric coefficients.  
Panel (a) shows two canonical omega particles, where one of them is equipped with a set of switches. When the orange switches are on and the magenta switches are off, the two particles are identical and exhibit identical magnetoelectric coupling of the omega type. However, when the orange switches are off and the magenta ones are on, the omega coupling coefficient of the right particle changes sign. Thus, the coupling effects due to the left and right particles cancel out, and, in total, there is no coupling. Panel (b) indicates a similar possibility for chiral mixtures where a proper setting of switches transforms a chiral structure into a racemic one.   

On the other hand, panels (c) and (d) demonstrate the concept of Tellegen and artificially moving meta-atoms as near-field coupled magnetized resonant ferrite spheres and properly shaped metal wires or strips (for microwave frequency operations)~\cite{Tellegen_exp,transparent,mirmoosaMTPRAF}. In this case, the nonreciprocal coupling can be switched on and off by using a switch in the bias circuit. Another way to realize artificial moving media is to employ equivalent transmission lines~\cite{Vehmas_2014}. Based on this method, nonreciprocal coupling is realized by using active electronic circuits acting as a gyrator. The coupling strength can be electronically tuned or switched on/off by regulating the circuit response. Yet another approach is based on the use of nonreciprocal coupling between two uniform antenna arrays~\cite{Caloz}. In this case, one of the arrays acts as a receiving array, whose received signal is passed to the other, transmitting array via a set of amplifiers. Amplifiers are nonreciprocal components passing the wave only in one direction. In Ref.~\cite{Caloz}, it was shown that this active metasurface is equivalent to a thin layer of an artificially moving medium. The artificial velocity can be simply regulated by changing the regime of the amplifiers or switching them off completely.

%%%%%%%%%%%%%%%%%%%%%%%%%%%%%%%%%%%%%%%%%%%%%%%%%%%%%%%%%%%%%%%%%%%%%%%%%%%%%%%%%%%%
%%%%%%%%%%%%%%%%%%%%%%%%%%%%%%%%%%%%%%%%%%%%%%%%%%%%%%%%%%%%%%%%%%%%%%%%%%%%%%%%%%%%
%%%%%%%%%%%%%%%%%%%%%%%%%%%%%%%%%% Section 4 %%%%%%%%%%%%%%%%%%%%%%%%%%%%%%%%%%%%%%%

\section{Discussion and Conclusions} 
\label{sec:CONCL}

%In contrast to a single spatial interface, the wave scattering was proved to be much more profound for a temporal interface between bianisotropic and magnetodielectric media. 

\begin{figure*}[h!]
\centerline
{\includegraphics[width=0.9\linewidth]{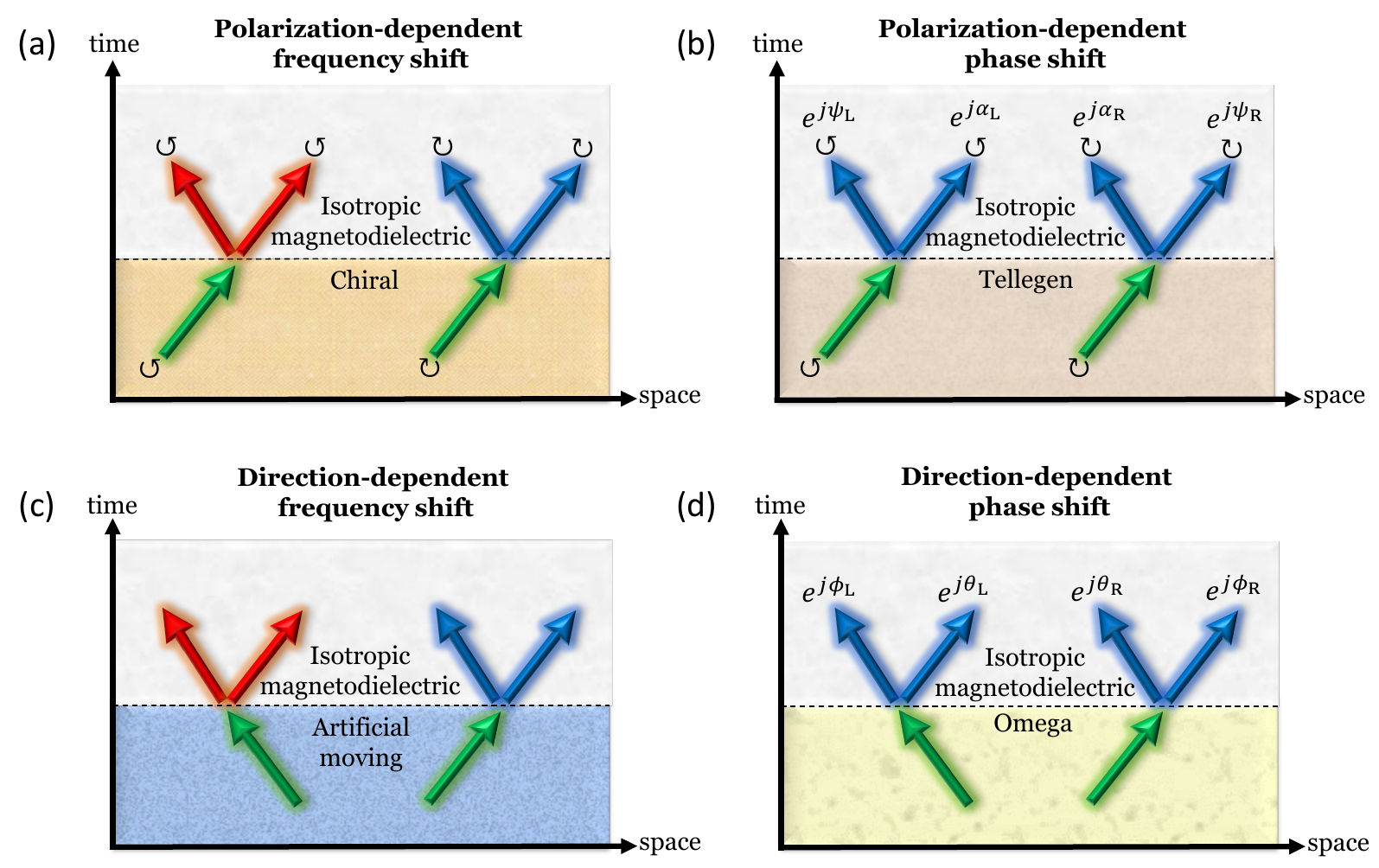}} 
\caption{A schematic illustration of different scattering phenomena at single temporal interfaces between bianisotropic media of different classes and a magnetodielectric. Different colors indicate different frequencies, rotating arrows indicate handedness of circular polarization, and the complex exponentials indicate the phase shifts taking place at time interfaces.}
\label{fig_time}
\end{figure*}

The full set of electromagnetic effects for plane waves at temporal interfaces between bianisotropic media and an isotropic magnetodielectric is graphically illustrated in Fig.~\ref{fig_time}. The illustrations in Figs.~\ref{fig_space} and \ref{fig_time} show comparative sets of phenomena at both space and time interfaces. A general feature of time interfaces is frequency conversion in contrast to propagation constant change at space interfaces, and this feature is common also to jumps of bianisotropic parameters. For a temporal interface, first, the frequencies of created waves and all the amplitudes of the transmitted and reflected waves depend strikingly on the magnetoelectric coupling coefficient value before the jump. Second, we have found directional-dependent and polarization-dependent angular frequency translations that complement the corresponding asymmetric effects at spatial interfaces, as listed in Table~\ref{tab:comp}. Regarding Tellegen and omega media, for both time and space interfaces, waves experience a polarization rotation and a phase shift, respectively. At time interfaces, these effects are accompanied by frequency translations. 

\begin{table*}[h!]
\center
\begin{tabular}{ |p{0.08\linewidth}||p{0.13\linewidth}||p{0.12\linewidth}||p{0.21\linewidth}||p{0.18\linewidth}|p{0.18\linewidth}| }
%\hline
%\multicolumn{3}{|c|}{Phenomena} \\
\hline
Class & Symmetry & Reciprocity & Phenomena & Temporal Interface & Spatial Interface \\
\hline 
\hline
Chiral & Symmetric & Reciprocal &  Polarization-dependent & Frequency shift & Wavelength shift\\ 
\hline
Moving & Antisymmetric & Nonreciprocal & Directional-dependent & Frequency shift & Wavelength shift \\
\hline
Tellegen & Symmetric & Nonreciprocal & Polarization-dependent & Phase shift & Phase shift \\
\hline 
Omega & Antisymmetric & Reciprocal & Directional-dependent & Phase shift & Phase shift \\
\hline
\end{tabular} 
\caption{Different scattering phenomena (in transmission and reflection) from single temporal and spatial interfaces between different classes of bianisotropic media and a magnetodielectric.}
\label{tab:comp}
\end{table*}

As was discussed in Section~\ref{ss_ms}, magnetoelectric parameters of the four classes of bianisotropic media define and control all possible symmetry breaking in response to waves illuminating the opposite sides of static uniaxial metasurfaces. Obviously, if the metasurface is static, the frequency of the incident waves is always preserved. In this study, we have found a possibility to break also the symmetry of frequency conversions at time inhomogeneities if the medium is bianisotropic. We expect that future studies of time-modulated bianisotropic metasurfaces will reveal that the identified direction-dependent phenomena at single interfaces will allow full control over the response of metasurfaces at all illuminations, not only in terms of the amplitude, phase, and polarization but also frequencies of the scattered waves. Furthermore, it is anticipated that further investigations of such sophisticated problems as bianisotropic temporal slabs, in which the magnetoelectric coupling can be turned on/off several times, will reveal even more drastic and multifaceted effects. Finally, we note the identified extreme phenomena of dramatic changes in the frequency and amplitude of waves, even up to the complete elimination of the electromagnetic field of a propagating wave for a specific value of one of the magnetoelectric coefficients.

%%%%%%%%%%%%%%%%%%%%%%%%%%%%%%%%%%%%%%%%%%%%%%%%%%%%%%%%%%   %%%%%%%%%%%%%%%%%%%%%%%%%%%%%%%%%%%%%%%%%%%%%%%%%%%%%%%%%%
%%%%%%%%%%%%%%%%%%%%%%%%%%%%%%%%%%%%%%%%%%%%%%%%%%%%%%%%%%

\section*{Acknowledgment}
This work was supported by the Research Council of Finland (Grants Nos.~330260, 336119, 354918, and 346518).

%%%%%%%%%%%%%%%%%%%%%%%%%%%%%%%%%%%%%%%%%%%%%%%%%%%%%%%%%%
%%%%%%%%%%%%%%%%%%%%%%%%%%%%%%%%%%%%%%%%%%%%%%%%%%%%%%%%%%
%%%%%%%%%%%%%%%%%%%%%%%%%%%%%%%%%%%%%%%%%%%%%%%%%%%%%%%%%%

\bibliographystyle{IEEEtran}
\bibliography{IEEEabrv,references}

\section*{Appendix} 
Let us assume  that a linearly polarized plane wave is propagating in a bianisotropic medium of one of the four classes, illuminating a spatial interface with  free space. The interface is at the plane $z=0$, and we study the case of normal incidence. The corresponding incident electric field is given by $\_E=E_0\exp(-j\beta z)\exp(j\omega t)\_a_x$. In the following, we briefly discuss the reflection and transmission phenomena separately for each class. 

\noindent{\textbf{Tellegen coupling:}} We write the incident electric field as a sum of fields associated with right-handed and left-handed circularly polarized plane waves. Equations~\eqref{ET:eq} and \eqref{HT:eq} describe such fields. Due to the existence of a spatial interface, we have reflected waves that propagate within the Tellegen medium and whose fields are expressed similarly to the incident right and left-handed components with a difference that $\beta$ changes sign ($\beta\rightarrow-\beta$). The reflection ($\Gamma$) and transmission ($\tau$) coefficients for right- and left-handed components are denoted as $\Gamma_{\/R}$, $\tau_{\/R}$ and $\Gamma_{\/L}$, $\tau_{\/L}$, respectively. It is worth noting that in contrast to the temporal interface, the angular frequency is conserved.
%{\color{red}, and the wavenumber in free space is converted.
%--- Do not understand: in uniform free space, of course the wavenumber is constant, but it changes at the interface.}
What we need to do is to impose the boundary conditions which are the continuity of tangential electric and magnetic fields at $z=0$. Thus, concerning the electric field, we have 
\begin{equation}
\frac{E_0}{2}+\Gamma_{{\/R},{\/L}}=\tau_{{\/R},{\/L}},\qquad 
\end{equation}
and regarding the magnetic field, we write   
\begin{equation}
{1\over2\mu_{\/F}}\Big({j\beta\over\omega}-{\chi_{\/F}\over c}\Big)E_0-{1\over\mu_{\/F}}\Big({j\beta\over\omega}+{\chi_{\/F}\over c}\Big)\Gamma_{\/R}={j\over\eta_0}\tau_{\/R},\qquad 
-{1\over2\mu_{\/F}}\Big({j\beta\over\omega}+{\chi_{\/F}\over c}\Big)E_0+{1\over\mu_{\/F}}\Big({j\beta\over\omega}-{\chi_{\/F}\over c}\Big)\Gamma_{\/L}=
-{j\over\eta_0}\tau_{\/L}. 
\end{equation}
Based on the above two equations, we derive the corresponding coefficients describing the reflection and transmission of the orthogonal circularly-polarized components. Assuming that the material parameters and the propagation constant $\beta$ are real-valued, we see that the transmission coefficients are complex conjugate of each other: 
\begin{equation}
\tau_{\/R}=\frac{\eta_0\sqrt{\mu_{\/F}\epsilon_{\/F}-\chi_{\/F}^2/c^2}}{\eta_0\sqrt{\mu_{\/F}\epsilon_{\/F}-\chi_{\/F}^2/c^2}+\mu_{\/F}-j\mu_0\chi_{\/F}}E_0,\qquad 
\tau_{\/L}=\frac{\eta_0\sqrt{\mu_{\/F}\epsilon_{\/F}-\chi_{\/F}^2/c^2}}{\eta_0\sqrt{\mu_{\/F}\epsilon_{\/F}-\chi_{\/F}^2/c^2}+\mu_{\/F}+j\mu_0\chi_{\/F}}E_0.
\end{equation} 
Consequently, due to this important feature,  the total transmitted field has both $x$ and $y$ components, which means that the linearly polarized incident field is rotated in the plane. The same conclusion is achieved for the reflected field. 

\noindent{\textbf{Chiral coupling:}} Equation~\eqref{eq:HMAGCHI} explicitly indicates that the amplitude of the magnetic field is not a function of the chirality parameter for both right-handed and left-handed circularly polarized waves. From this point of view, these waves have the same coefficients of transmission and reflection. The important result is that if the total incident electric field at the spatial interface has only one component such as an $x$-component (as we initially wrote at the beginning of the section), the total transmitted field possesses also only this component. However, notice that the reflected field vector rotates in the plane as it continuously propagates back in the chiral medium. 

\noindent{\textbf{Moving coupling:}} In this case,  the medium is uniaxial, and we consider the case where the axis is normal to the interface. As we showed in Subsection~\ref{subsec:DDPMO}, the corresponding magnetic field amplitude does not depend on the artificial velocity parameter regardless of the propagation direction. This is similar to the chiral medium explained above. Thus, for both positive ($+z$) and negative ($-z$) directions, the reflection and transmission coefficients are also independent of the magnetoelectric coupling parameter. 

\noindent{\textbf{Omega coupling:}} We eventually describe the interface phenomena for the case in which we have a spatial interface between an omega medium and free space. According to Subsection~\ref{subsec:DDPMO}, the propagation constant is an even function with respect to the omega parameter. Hence, the sign of the omega parameter does not affect the propagation constant. However, the magnetic field is related to the omega parameter so  that its sign has an impact on the phase. Indeed, after some algebraic manipulations, we readily derive that    
\begin{equation}
\tau=\frac{2\eta_0\sqrt{\mu_{\/F}\epsilon_{\/F}-\Omega_{\/F}^2/c^2}}{\eta_0\sqrt{\mu_{\/F}\epsilon_{\/F}-\Omega_{\/F}^2/c^2}+\mu_{\/F}+j\mu_0\Omega_{\/F}}E_0,
\end{equation}
which means that changing the sign of $\Omega_{\/F}$ changes the phase of the transmission coefficient (or the reflection coefficient). Specifically, if the material parameters are real values, based on the above expression, we conclude that when $\Omega_{\/F}$ is changed to $-\Omega_{\/F}$, the transmission coefficient becomes complex conjugate of the original value.

%%%%%%%%%%%%%%%%%%%%%%%%%%%%%%%%%%%%%%%%%%%%%%%%%%%%%%%%%
%%%%%%%%%%%%%%%%%%%%%%%%%%%%%%%%%%%%%%%%%%%%%%%%%%%%%%%%%
%%%%%%%%%%%%%%%%%%%%%%%%%%%%%%%%%%%%%%%%%%%%%%%%%%%%%%%%%

%\section*{Author contributions}

%\section*{Competing interests}
%All authors have no competing interests.

%\section*{Data and materials availability}
%All data is available in the manuscript. 

\end{document}